\documentclass[aps,preprint,epsfig,rotate]{revtex4}
\usepackage{graphicx}
\usepackage{bm}
\usepackage{epsfig}

\input epsf
\begin{document}
\title{On the emission of the fast $\delta-$electrons during nuclear $\beta^{-}$-decay in few-electron atoms}

\author{Alexei M. Frolov}
\email[E--mail address: ]{afrolov@uwo.ca}

\affiliation{Department of Applied Mathematics \\
 University of Western Ontario, London, Ontario N6H 5B7, Canada}

\date{\today}

\begin{abstract}

We discuss a possibility to observe fast secondary electrons ($\delta-$electrons) during nuclear $\beta^{-}$-decay in few-electron 
atoms/ions. Formulas for the corresponding probabilities and velocity spectrum of the $\delta-$electrons are derived in the closed 
analytical forms.

\noindent 
PACS number(s): 23.40.-s, 12.20.Ds and 14.60.Cd


\end{abstract}

\maketitle
\newpage

As follows from numerous experiments in a relatively large number of cases the nuclear $\beta^{-}$-decay in few-electron atoms leads to the `additional' ionization of the 
final ion. In general, such a process in the $\beta^{-}$-decaying atom X is described by the following equation (see, e.g., \cite{Fro1}):
\begin{equation}
  X \rightarrow Y^{2+} + e^{-} + e^{-}(\beta) + \overline{\nu} \label{eq1}
\end{equation}
where the symbols $X$ and $Y$ designate two different chemical elements (isotopes) with almost equal masses. The sybmols X and Y in Eq.(\ref{eq1}) are used to designate both 
atoms/ions and the corresponding atomic nuclei. If $Q$ is the electric charge of the incident nucleus $X$, then the nuclear charge of the final nucleus $Y$ is $Q + 1$. Below, 
the electric charge of the incident nucleus ($Q$) is designated by the notation $Q_1$, while the electric charge of the final nucleus is denoted by the notation $Q_2 (= Q + 
1)$. In Eq.(\ref{eq1}) the notation $e^{-}$ stands for the secondary (or slow) electron formed in the unbound spectrum during the decay, Eq.(\ref{eq1}). Also, in Eq.(\ref{eq1}) 
the notation $e^{-}(\beta)$ designates the fast $\beta^{-}-$electron and $\overline{\nu}$ denotes the electron's anti-neutrino. In actual atoms/ions, the overall probability of 
the process, Eq.(\ref{eq1}), varies between 2.5 \% and 35 \% - 50 \% \cite{Fro05} - \cite{PRC2}. Our method used in \cite{Fro05} to evaluate the final state probabilities is 
based on the sudden approximation \cite{LLQ}, \cite{MigK} which means that the velocities of atomic electrons $v_{a}$, including electrons emitted during the process, 
Eq.(\ref{eq1}), are significantly smaller (100 - 1000 times smaller) than the velocity $v_{\beta}$ of the $\beta^{-}$-particle (or $\beta^{-}$-electron). For atoms, ions and 
molecules this means that the electron density distribution cannot change during nuclear $\beta^{-}$-decay. This substantially simplifies numerical evaluations of the final 
state probabilities. 

In this study we apply sudden approximation to study another important process which has smaller overall probability, but it is of a great interest in a number of experiments 
and applications. The equation of this process is
\begin{equation}
  X \rightarrow Y^{2+} + e^{-}(\delta) + e^{-}(\beta) + \overline{\nu} \label{eq2}
\end{equation}
where $e^{-}(\delta)$ is the fast scondary electron emitted and accelerated to relatively large velocities during the nuclear $\beta^{-}$-decay. In a number of books and textbooks 
such fast electrons are called the $\delta-$electrons. The velocity of these $\delta-$electron is less, but always comparable with the velocity of the $\beta-$electron. Sudden 
acceleration to large and very large velocities proceeds when the emitted $\beta^{-}-$electron interact with slow electron from Eq.(\ref{eq1}) and transfers to that electron a 
substantial part of its incident momentum. It is clear that the probability of such a process is small. In the lowest-order approximation such a probability is evaluated as $P 
\approx \alpha^4 P_e$, where $P_e$ is the probability of the free-electron emission in the process, Eq.(\ref{eq1}) and $\alpha = \frac{e^2}{\hbar c} \approx \frac{1}{137}$ is the 
dimensionless fine-structure constant. More accurate evaluation leads to the formula which contain an additional factor which increases the numerical value of $P$. The main goal 
of this study is to obtain the more accurate formulas to evaluate the overall probability of emission of the fast $\delta-$electrons during $\beta^{-}$ decay in few-electron atoms 
and ions. In addition to this, we also want to predict some properties of the fast $\delta-$elelctrons, Eq.(\ref{eq2}), e.g., their energy distribution.      

It is clear that the emission of the fast $\delta-$electron during nuclear $\beta^{-}$ decay in few-electron atoms/ions is a pure QED process. The source of the fast $\delta-$electrons 
is the electromagnetic interaction between the fast $\beta^{-}$-particle (electron) and atomic electrons. Briefly, we need to consider the non-elastic electron-electron scattering 
where one of the incident electrons was very fast, while another electron was at rest. First, consider the case when $\beta^{-}$-decaying atom has one bound electron only. In actual 
experiments this case corresponds to the tritium atom ${}^{3}$H. In this case we can write the following formula known from QED (see, e.g, \cite{Grein}, \cite{AB}) for the 
cross-section of the $(e^{-},e^{-})$-scattering: 
\begin{eqnarray}
 d\sigma = 2 \pi \alpha^4 a^{2}_{0} \Bigl(\frac{\gamma_1 + 1}{\gamma^{2}_1 v^{4}_1}\Bigr) \Bigl[ \frac{4}{\tau^{4}} - \frac{3}{\tau^{2}} + \frac{(\gamma^{2}_1 - 1)^{2}}{4 \gamma^{2}_1} 
 \Bigl(1 + \frac{4}{\tau^{2}}\Bigr)\Bigr] \sin\theta d\theta  \label{eq3}
\end{eqnarray}  
where $\gamma_1 = \frac{E_1}{m c^2}$ is the Lorentz $\gamma-$factor of the $\beta-$electron, $\alpha = \frac{e^{2}}{\hbar c} \approx 7.2973525664 \cdot 10^{-3}$ is the dimensionless 
fine-structure constant and $a_0 = \frac{\hbar^2}{m_e e^2} \approx 5.2917721092(17) \cdot 10^{-9}$ $cm$ is the Bohr radius. In Eq.(\ref{eq3}) the factor $\tau^{2}$ is 
\begin{eqnarray}
  \tau^{2} = 1 - \Bigl[\frac{2 - (\gamma_1 + 3) \sin^{2}\theta}{2 + (\gamma_1 - 1) \sin^{2}\theta}\Bigr]^{2} \label{eq4}
\end{eqnarray}
where $\theta$ is the azimuthal angle, or scattering angle, if we take into account the cylindrycal symmetry of the scattering problem formulated above. The formulas, Eqs.(\ref{eq3}) 
and (\ref{eq4}) allow one to determine the angular distribution (or $\theta-$distribution) of the fast secondary electrons ($\delta-$electrons) emitted during the nuclear $\beta^{-}$ 
decay of the one-electron tritium atom ${}^{3}$H, or an arbitrary one-electron ion. 

Now, by using the formulas, Eqs.(\ref{eq3}) and (\ref{eq4}), we can represent the differential cross-section, Eq.(\ref{eq3}) in a different form. First, let us introduce the dimensionless 
parameter $\Delta = \frac{\gamma_2 - 1}{\gamma_1 - 1}$ ($0 \le \Delta \le 1$) which equals to the ratio of energies of the $\delta-$ and $\beta-$electrons emitted in the process, 
Eq.(\ref{eq2}). The parameters $\gamma_1$ and $\gamma_2$ in this formula are the $\gamma-$factors of the $\beta$- and $\delta$-electrons, respectively. Here and everywhere below, we 
assume that the $\delta-$electron from Eq.(\ref{eq3}) was at rest before nuclear $\beta^{-}$-decay (atomic velocities are negligible in comparison to $v_{\beta}$). By using a 
standard set of the conservation laws one can transform the formula for the parameter $\Delta$ to the following (equivalent) form(s): $\Delta = \frac{\gamma_1 - 
\gamma^{\prime}_1}{\gamma_1 - 1} = \frac12 (1 - \tau)$, where $\gamma^{\prime}_1$ is the Lorentz $\gamma-$factor of the $\beta^{-}$ electron at the final stage (after the scattering)
and parameter $\tau$ is defined by Eq.(\ref{eq4}). Substitution of the last expression in Eq.(\ref{eq3}) leads to the following formula for the differential cross-section:
\begin{eqnarray}
  d\sigma &=& \frac{2 \pi \alpha^4 a^{2}_{0} \gamma^{2}_1}{(\gamma^{2}_1 - 1) (\gamma_1 - 1)} \cdot \frac{d\Delta}{\Delta^2 (1 - \Delta)^2} \Bigl\{ 1 - \Bigl[ 3 - \Bigl(\frac{\gamma_1 
 - 1}{\gamma_1}\Bigr)^2 \Bigr] \Delta ( 1 - \Delta) \nonumber \\
 &+& \Bigl(\frac{\gamma_1 - 1}{\gamma_1}\Bigr)^2 \Delta^2 (1 - \Delta)^2 \Bigr\} \label{eq5}
\end{eqnarray} 
This formula gives the energy distribution for the secondary electrons (or delta-electrons) from the process, Eq.(\ref{eq2}). For small $\Delta$ one finds from Eq.(\ref{eq5}): $d\sigma = 
\frac{2 \pi \alpha^4 a^{2}_{0} \gamma^{2}_{1}}{(\gamma^{2}_1 - 1) (\gamma_1 - 1)} \cdot \frac{d\Delta}{\Delta^2}$. 

The formula, Eq.(\ref{eq5}), can be used to determine the final state probabities of the emission of the fast secondary electrons (or $\delta-$electrons) during nuclear $\beta^{-}$-decay 
in few-electron atoms. Being renormalized in a special way the same formula can be used to describe the energy spectrum of the fast $\delta-$electrons emitted in the process, Eq.(\ref{eq2}). 
However, in actual applications to few-electron atoms and ions the formula, Eq.(\ref{eq5}), must include an additional factor which represents a correction to the finite electron-electron 
distances of the electron-electron scattering in any actual $\beta^{-}$-decaying atom. Derivation of the formula, Eq.(\ref{eq5}), in QED is based on the assumption that the incident 
electron-electron distance is infinite. Evaluation of the finite-distance correction can be based on the fact that in actual atoms the distance $2 a_{0}$ can be considered as `infinte' 
distance, while actual atomic electrons are located at the distance $\langle r_{eN} \rangle \ll 2 a_0$ from the atomic nucleus which emits the fast $\beta^{-}$ electron. By comparing the 
corresponding solid angles in both cases one finds that the additional factor which must be included in the formula, Eq.(\ref{eq5}), is written in the form $\langle \frac{8}{r^{2}_{eN}} 
\rangle$. The final form of the equations, Eq.(\ref{eq5}), is
\begin{eqnarray}
  d\sigma &=& \zeta \frac{16 N_e \pi \alpha^4 a^{2}_{0} \gamma^{2}_1}{(\gamma^{2}_1 - 1) (\gamma_1 - 1)} \cdot \langle \frac{a^{2}_0}{r^{2}_{eN}} \rangle \cdot \frac{d\Delta}{\Delta^2 (1 
  - \Delta)^2} \Bigl\{ 1 - \Bigl[ 3 - \Bigl(\frac{\gamma_1 - 1}{\gamma_1}\Bigr)^2 \Bigr] \Delta ( 1 - \Delta) \nonumber \\
  &+& \Bigl(\frac{\gamma_1 - 1}{\gamma_1}\Bigr)^2 \Delta^2 (1 - \Delta)^2 \Bigr\} \label{eq55}
\end{eqnarray} 
where the $N_e$ is the total number of bound electrons in the incident atom/ion, while $\langle \frac{a^{2}_0}{r^{2}_{eN}} \rangle$ is the atomic expectation value which must be computed 
with the use of accurate atomic wave functions. Additional factor $\zeta$ in the formula, Eq.(\ref{eq55}), is used to produce the `best' approximation of the formula, Eq.(\ref{eq55}) to the 
experimental results known for the fast $\delta-$electrons. Now, the expression, Eq.(\ref{eq55}), can be used for actual atoms and ions. For instance, for the beta-decaying two-electron 
Na$^{9+}$ ion (in its ground $1^1S-$state) we have $N_e = 2$ and $\langle \frac{1}{r^{2}_{eN}} \rangle \approx$ 230.2388870075436700 $\approx$ 230.239 $a.u.$ \cite{Fro2015} ($a_0 = 1$ in 
$a.u.$). Therefore, the coefficient $16 N_e \pi \alpha^4 a^{2}_{0} \cdot \langle \frac{1}{r^{2}_{eN}} \rangle$ in the right-hand side of this equation is 1.83799$\cdot 10^{-21}$ $cm^{2}$ (at 
$\zeta = 1$). For analogous Co$^{25+}$ ion (in its ground $1^1S-$state) we have $\langle \frac{1}{r^{2}_{eN}} \rangle \approx$ 1428.872976503642391 $\approx$ 1428.873 $a.u.$ \cite{Fro2015}, 
and therefore, this coefficient is 1.140662$\cdot 10^{-20}$ $cm^{2}$. For the four-electron Be atom in its ground $2^1S-$state we have found that 2.299034$\cdot 10^{-22}$. These examples 
indicate clearly that the factor $16 N_e \pi a^{2}_{0} \cdot \langle \frac{1}{r^{2}_{eN}} \rangle$ in Eq.(\ref{eq55}) is a relatively large value (even for light atoms and ions). Finally, it 
is possible to say that the overall probability to observe fast $\delta-$electrons is $\approx \alpha^3$ (or even $\approx \alpha^2$). In turn, this unambiguously leads to the conclusion that 
the emission of the fast secondary electrons (or $\delta-$electrons) can be observed in modern experiments. 

For actual $\beta^{-}-$decaying atoms the formulas, Eqs.(\ref{eq4}), (\ref{eq5}) and (\ref{eq55}), must be integrated (over the variable $\gamma_1$) with the actual spectrum of the emitted 
$\beta^{-}$ electrons \cite{Fermi} (see also \cite{Long} - \cite{Mei}, \cite{Konop} and references therein). For instance, the explicit formula for the probability to detect the final 
$\delta-$electron with the Lorentz $\gamma-$factor $\gamma_2$ is
\begin{eqnarray}
  P(\gamma_2) = \int_{1}^{\gamma_{max}} \Bigl(\frac{d\sigma}{d \Delta}\Bigr) \Bigl(\frac{d\Delta}{d\gamma_1}\Bigr) S_{\beta}(\gamma_1) d\gamma_1 = (\gamma_2 - 1) \int_{1}^{\gamma_{max}} 
  \Bigl(\frac{d\sigma}{d \Delta}\Bigr) S_{\beta}(\gamma_1) \frac{d\gamma_1}{(\gamma_1 - 1)^2} \label{eq7}
\end{eqnarray} 
which can formally be considered as an integral transformation of the spectra of $\beta^{-}$ electrons (primary fast electrons). The kernel of this integral transformation equals 
$\Bigl(\frac{d\sigma}{d \Delta}\Bigr) \Bigl(\frac{d\Delta}{d\gamma_1}\Bigr) = \frac{1}{(\gamma_1 - 1)^2} \Bigl(\frac{d\sigma}{d \Delta}\Bigr)$. In this equation the derivative 
$\frac{d\sigma}{d \Delta}$ is the differential cross-section defined by Eq.(\ref{eq55}), while another derivative $\frac{d\Delta}{d\gamma_1}$ is determined from the definition of $\Delta$ 
(see above). The spectral function $S_{\beta}(\gamma_1)$ takes the form
\begin{eqnarray}
  S_{\beta}(\gamma_1) d\gamma &=& {\cal C}_{\gamma} \cdot F(Q + 1, (\gamma - 1) m_e c^2) \cdot \Bigl[ \frac{\Delta E^{\prime} +  m_e c^{2}}{m_e c^{2}} - \gamma \Bigr]^2 
  (\gamma^2 - 1)^{\frac12} \cdot \gamma d\gamma \label{eq65} \\
  &=& {\cal C}^{\prime}_{\gamma} \cdot F(Q + 1, \gamma - 1) \cdot \Bigl[ \frac{\Delta E^{\prime}}{m_e c^{2}} - \gamma \Bigr]^2 (\gamma^2 - 1)^{\frac12} \cdot \gamma d\gamma \nonumber 
\end{eqnarray} 
This expression almost excatly coincides with the formula, Eq.(210), derived in \cite{Bethe}, which, however, contains no Fermi function introduced in \cite{Fermi}. In general, the assumption 
that $F(Q + 1, \gamma - 1) = 1$ works well for light atoms, but for intermedium ($Q \ge 40$) and heavy ($Q \ge 75$) atoms the actual Fermi factor is needed. Accurate numerical evaluations of 
the Fermi function have been performed in a large number of papers (see, e.g., \cite{Long}, \cite{Mei}). Experimental energy spectra of the emited primary $\beta^{-}$ electrons can be found, 
e.g., in \cite{Cook} and \cite{Neary}, where the $\beta^{-}$ decays of the ${}^{64}$Cu and ${}^{210}$Bi atoms were studied in detail. As follows from Eq.(\ref{eq65}), the upper limit in the
formula, Eq.(\ref{eq65}), equals $\gamma_{max} = \frac{\Delta E^{\prime}}{m_e c^{2}}$. The velocity/energy spectra of the fast $\delta-$electrons are determined from the equations which are
similar to Eq.(\ref{eq65}).   

To conclude our analysis let us note that emission of the fast secondary electrons can also be observed during nuclear $\beta^{+}$-decay in few-electron atoms. Most of the formulas derived  
above can also be applied to description of such process. However, there are some differences which are quite obvious. A detalied analysis of this process will be performed in our next study.


\begin{thebibliography}{01}

\bibitem{Fro1} A.M. Frolov and D.M. Wardlaw, \textit{On the spectrum of secondary electrons emitted during nuclear $\beta^{-}$-decay in few-electron atoms}, 
ArXiv: 1511.06475v1 [phys.at.-ph.] (2015).

\bibitem{Fro05} A.M. Frolov and J.D. Talman, Phys. Rev. A {\bf 72}, 022511 (2005) [see also: A.M. Frolov and M.B. Ruiz, Phys. Rev. A {\bf 82}, 042511 (2010);
Adv. Quant. Chem. {\bf 67}, 267 (2013)].

\bibitem{PRC1} F. Simkovic, R. Dvornicky and A. Faessler, Phys. Rev. C {\bf 77}, 055502 (2008). 

\bibitem{PRC2} N. Doss, J. Tennyson, A. Saenz and S. Jonsell, Phys. Rev. C {\bf 73}, 025502 (2006). 

\bibitem{LLQ} L.D. Landau and E.M. Lifshitz, {\it Quantum Mechanics: Non-Relativistic Theory}, (3rd. ed. Pergamon Press, New York (1976)), Chpt. VI. 

\bibitem{MigK} A.B. Migdal and V. Krainov, {\it Approximation Methods in Quantum Mechanics}, (W.A. Benjamin, New York (1969)). 

\bibitem{Grein} W. Greiner and J. Reinhardt, {\it Quantum Electrodynamics}, (4th edn., Springer Verlag, Berlin, 2009).

\bibitem{AB} A.I. Akhiezer and V.B. Beresteskii, {\it Quantum Electrodynamics}, (4th ed., Science, Moscow (1981)), Chpt. 4 and 5.

\bibitem{Fro2015} A.M. Frolov, Chem. Phys. Lett. {\bf 638}, 108 - 115 (2015).

\bibitem{Fermi} E. Fermi, Zeits. f\"{u}r Physik {\bf 88}, 172 (1934).

\bibitem{Long} C. Longmire and H. Brown, Phys. Rev. {\bf 75}, 264 (1949).

\bibitem{Hall} H. Hall, Phys. Rev. {\bf 79}, 745 (1950).

\bibitem{Mei} J.Y. Mei, Phys. Rev. {\bf 81}, 287 (1951).

\bibitem{Konop} E.J. Konopinski, Rev. Mod. Phys., {\bf 15}, 209 (1943).

\bibitem{Cook} L.S. Cook and L.M. Langer, Phys. Rev., {\bf 73}, 601 (1948).

\bibitem{Neary} G. J. Neary, Roy. Phys. Soc. (London), {\bf A175}, 71 (1940).

\bibitem{Bethe} H.A. Bethe and R.F. Bacher, Rev. Mod. Phys., {\bf 8}, 82 (1935).

\end{thebibliography}
\end{document}